\documentclass[prc,aps,
twocolumn,
tightenlines,
preprintnumbers,
superscriptaddress,
a4paper,nofootinbib,10pt]{revtex4-2}
\usepackage{adjustbox}
\usepackage{graphicx}
\usepackage{tikz}
\usepackage{epsf}
\usepackage{amsmath,amssymb}             
\usepackage{cases}
\usepackage{amsfonts}
\usepackage{mathrsfs}
\usepackage{xcolor}
\usepackage{cancel}
\usepackage{nicefrac}
\usepackage{array}
\usepackage{tabularx}     
\usepackage[normalem]{ulem}
\usepackage{adjustbox}
\usepackage{makecell}
\usepackage{graphicx}
\usepackage{booktabs}
\usepackage{times}
\usepackage{siunitx}
\usepackage[colorlinks,citecolor=blue,linktoc=all,linkcolor=cyan,urlcolor=blue]{hyperref}

\usepackage{enumitem}

\newcolumntype{L}[1]{>{\raggedright\let\newline\\\arraybackslash\hspace{0pt}}m{#1}}
\newcolumntype{C}[1]{>{\centering\let\newline\\\arraybackslash\hspace{0pt}}m{#1}}
\newcolumntype{R}[1]{>{\raggedleft\let\newline\\\arraybackslash\hspace{0pt}}m{#1}}

\usepackage{comment}

\usepackage[colorlinks,citecolor=blue,linktoc=all,linkcolor=cyan]{hyperref} 

\def\beq{\begin{equation}}
\def\eeq{\end{equation}}
\def\bea{\begin{eqnarray}}
\def\eea{\end{eqnarray}}
\def\beqa{\begin{equation}\begin{array}{l}}
\def\eeqa{\end{array}\end{equation}}
\def\eqlab#1{\label{eq:#1}}

\def\seclab#1{\label{sec:#1}}
\def\eref#1{(\ref{eq:#1})}
\def\Eqref#1{Eq.~(\ref{eq:#1})}

\def\Figref#1{Fig.~\ref{fig:#1}}

\def\secref#1{Section~\ref{sec:#1}}

\def\barr{\left(\begin{array}{c}}
\def\earr{\end{array}\right)}
\def\bmat{\left(\begin{array}{cc}}
\def\emat{\end{array}\right)}
\def\al{\alpha}

\def\ga{\gamma}

\def\dd{{\rm d}}

\def\nn{\nonumber}

\usepackage{tikz}
\usepackage[compat=1.1.0]{tikz-feynman}
\usetikzlibrary{arrows.meta,shapes.misc}
\tikzset{
  thickwiggle/.style={
    decorate,
    decoration={snake, amplitude=6pt, segment length=15pt},
    draw,
    ultra thick
  }
}
\definecolor{myblobcolor}{RGB}{70,130,230}
\definecolor{LbLBlobColor}{RGB}{255, 157, 0}
\definecolor{VVCSspin1}{RGB}{80, 200, 120}
\newcolumntype{Y}{>{\raggedright\arraybackslash}X}

\usepackage{booktabs}          
\usepackage{tabularx}          
\usepackage{siunitx}           
\usepackage{threeparttable}    
\def\3d{3-D}

\newcommand{\twodots}{\mathinner {\ldotp \ldotp}}

\definecolor{olive}{HTML}{668000}
\definecolor{lightolive}{HTML}{CCFF00}
\definecolor{myorange}{HTML}{FF9900}
\definecolor{mylblue}{HTML}{8080FF}
\definecolor{mydgreen}{HTML}{008000}

\usepackage{array} 

\newif\iftest
\begin{document}
\preprint{MITP/26-XXX}

\title{Combined Analysis of Lattice QCD and Experimental Data on the Pion Transition Form Factor} 

\author{Franziska Hagelstein}
\affiliation{Institut f\"ur Kernphysik,
 Johannes Gutenberg-Universit\"at  Mainz,  D-55128 Mainz, Germany}
   \affiliation{PRISMA$^{++}$ Cluster of Excellence,
 Johannes Gutenberg-Universit\"at  Mainz,  D-55128 Mainz, Germany}
\affiliation{PSI Center for Neutron and Muon Sciences, CH-5232 Villigen PSI, Switzerland}
 \author{Danaheb Naomi Navarro Dur\'an}
\affiliation{Institut f\"ur Kernphysik,
 Johannes Gutenberg-Universit\"at  Mainz,  D-55128 Mainz, Germany}
\author{\\Timon Esser}
\affiliation{Institut f\"ur Kernphysik,
 Johannes Gutenberg-Universit\"at  Mainz,  D-55128 Mainz, Germany}
 \affiliation{PRISMA$^{++}$ Cluster of Excellence,
 Johannes Gutenberg-Universit\"at  Mainz,  D-55128 Mainz, Germany}
\author{Vadim Lensky}
\affiliation{Institut f\"ur Kernphysik,
 Johannes Gutenberg-Universit\"at  Mainz,  D-55128 Mainz, Germany}
 \author{Sotiris Pitelis}
\affiliation{Institut f\"ur Kernphysik,
 Johannes Gutenberg-Universit\"at  Mainz,  D-55128 Mainz, Germany}
 \author{Vladyslava Sharkovska}
\affiliation{Institut f\"ur Kernphysik,
 Johannes Gutenberg-Universit\"at  Mainz,  D-55128 Mainz, Germany}
\affiliation{PSI Center for Neutron and Muon Sciences, CH-5232 Villigen PSI, Switzerland}
\affiliation{Department of Physics, University of Zurich, CH-8057 Zurich, Switzerland}

\date{\today}

\begin{abstract}
The evaluation of the hadronic light-by-light scattering contribution to the muon anomalous magnetic moment requires precise knowledge of the pion transition form factor (TFF). In this work, we present a feasibility study for a combined analysis of lattice QCD (LQCD) and experimental data. Our methodology is driven by the goal of combining complementary datasets to leverage their respective kinematic advantages: while LQCD provides robust predictions for the doubly-virtual TFF, $e^+e^-$ scattering experiments offer high-precision singly-virtual measurements up to large momentum transfers. To ensure a statistically rigorous combination, we implement a global one-stage fitting approach based on the modified $z$-expansion, utilizing synthetic jackknife replicate sampling and a normalized $\chi^2$ weighting scheme. We demonstrate that the inclusion of experimental data substantially tightens the constraints on the pion TFF, yielding up to a factor of three reduction in uncertainty in the singly-virtual limit. In contrast, the uncertainty of the resulting pion-pole contribution to the muon $g-2$ improves by a factor of $1.5$. This more modest improvement reflects the fact that the $g-2$ integral is heavily dominated by the low-$Q^2$ region, which is already well constrained by physical normalization constraints.
\end{abstract}

\keywords{Pseudoscalar transition form factor; Pion-pole contribution; Hadronic light-by-light scattering; Muon anomalous magnetic moment}
\maketitle
\tableofcontents

\section{The pion transition form factor: Motivation for a combined analysis}

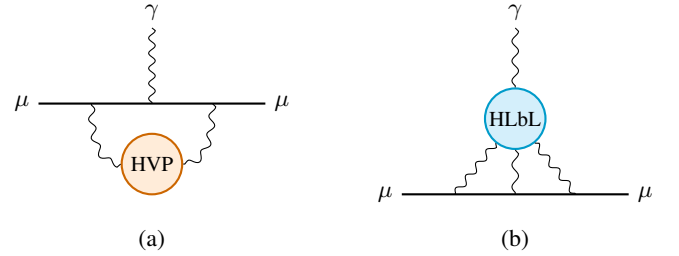
\begin{figure}[t]
    \centering

    \begin{tikzpicture}[baseline=0]
      \begin{feynman}
        \coordinate (mu1) at (-1.5, 0);
        \coordinate (v1)  at (-0.8, 0);
        \coordinate (v2)  at (0, 0);
        \coordinate (v3)  at (0.8, 0);
        \coordinate (mu2) at (1.5, 0);
        
        \node[circle, fill=orange!15, draw=orange!80!black, thick, minimum size=0.8cm, inner sep=0pt] 
          (hvp) at (0, -0.8) {\footnotesize HVP};
          
        \coordinate (gamma) at (0, 1.0);
        
        \draw[thick] (mu1) node[left] {\(\mu\)} -- (mu2) node[right] {\(\mu\)};
        
        \draw[photon] (v2) -- (gamma) node[above] {\(\gamma\)};
        
        \draw[photon] (v1) to[out=-90, in=180] (hvp);
        \draw[photon] (hvp) to[out=0, in=-90] (v3);
      \end{feynman}
      
      \node at (0, -1.8) {(a)};
    \end{tikzpicture}
    \hfill
    \begin{tikzpicture}[baseline=0]
      \begin{feynman}
        \coordinate (mu1) at (-1.5, -1.2);
        \coordinate (v1)  at (-0.8, -1.2);
        \coordinate (v2)  at (0, -1.2);
        \coordinate (v3)  at (0.8, -1.2);
        \coordinate (mu2) at (1.5, -1.2);
        
        \node[circle, fill=cyan!15, draw=cyan!80!black, thick, minimum size=0.8cm, inner sep=0pt] 
          (hadr) at (0, -0.2) {\footnotesize HLbL};
          
        \coordinate (gamma) at (0, 1.0);
        
        \draw[thick] (mu1) node[left] {\(\mu\)} -- (mu2) node[right] {\(\mu\)};
        
        \draw[photon] (v1) -- (hadr);
        \draw[photon] (v2) -- (hadr);
        \draw[photon] (v3) -- (hadr);
        
        \draw[photon] (hadr) -- (gamma) node[above] {\(\gamma\)};
      \end{feynman}
      
      \node at (0, -1.8) {(b)};
    \end{tikzpicture}%
    \caption{Hadronic contributions to $a_\mu$: (a) Hadronic vacuum polarization (HVP); (b) Hadronic light-by-light scattering (HLbL).}
    \label{fig:hadronic_g_minus_2}
\end{figure}

The muon anomalous magnetic moment $a_\mu\equiv\frac12 (g-2)_\mu$ provides a stringent precision test of the Standard Model of Particle Physics (SM). Presently, this test is limited by the SM theory prediction, as published by the ``Muon g-2 Theory Initiative'' (WP25) \cite{Aliberti:2025beg}. Comparing the experimental value $a_\mu^\text{exp}$ \cite{Muong-2:2025xyk} (world average of the Fermilab E989 and Brookhaven E821 experiments)  to the SM prediction $a_\mu^\text{SM}$, which is more than four times less precise, no discrepancy and, thus, no indication of New Physics is observed:
\begin{subequations}
\label{amuSM}
\begin{eqnarray}
a_\mu^\text{exp}&=&116\,592\,071.5(14.5)\times 10^{-11},\\
a_\mu^\text{SM, WP25}&=&116\,592\,033(62)\times 10^{-11},\\
a_\mu^\text{exp}-a_\mu^\text{SM, WP25}
&=&38(63)\times 10^{-11}.
\end{eqnarray}
\end{subequations}
This, however, has not always been the case. The uncertainty of the SM prediction is dominated by the hadronic contributions. Since the previous community consensus (WP20) \cite{Aoyama:2020ynm}, the methodology of their evaluation has changed in various aspects, thereby  mending a long-standing discrepancy between the SM prediction and the experimental value of $a_\mu$. 

The larger hadronic vacuum polarization (HVP) contribution, shown in \Figref{hadronic_g_minus_2} (a), increased by more than $3\,\sigma$. It is now based on the average of lattice QCD (LQCD)
predictions for the leading-order (LO) HVP \cite{RBC:2018dos,Giusti:2019xct,Borsanyi:2020mff,Lehner:2020crt,Wang:2022lkq,Aubin:2022hgm,Ce:2022kxy,ExtendedTwistedMass:2022jpw,RBC:2023pvn,Kuberski:2024bcj,Boccaletti:2024guq,Spiegel:2024dec,RBC:2024fic,Djukanovic:2024cmq,ExtendedTwistedMass:2024nyi,MILC:2024ryz,FermilabLatticeHPQCD:2024ppc}, instead of the classical data-driven dispersive evaluations (KNTW \cite{Keshavarzi:2018mgv,Keshavarzi:2024wow}, DHMZ \cite{Davier:2019can}). Empirical input for the latter from measurements of $e^+e^-$ annihilation into hadrons is presently under critical review. While previous smaller tensions between BaBar and KLOE could still be combined into a meaningful average with inflated uncertainties \cite{Aoyama:2020ynm}, the recent CMD-3 experiment \cite{CMD-3:2023alj,CMD-3:2023rfe} disagrees with older experiments (BaBar, BESIII, CMD-2, KLOE, SND) in the important $\pi^+\pi^-$ region. On the theoretical side, the ``RadioMonteCarlo 2 Working Group'' has reviewed and compared the available Monte Carlo generators \cite{Aliberti:2024fpq} and aims to provide state-of-the-art radiative corrections for $e^+e^-$ scattering at next-to-next-to-leading order (NNLO).  Hadronic $\tau$ decays can provide further empirical constraints. The study of ``window'' quantities in isolated kinematic regions (short, intermediate, and long distance) allows for a more detailed comparison between the discrepant LQCD and data-driven predictions. In the future, the MUonE experiment will provide an independent determination of the HVP contribution based on a measurement of the hadronic contribution to the running of the electromagnetic coupling constant $\Delta \al_\mathrm{had}(t)$.

The hadronic light-by-light scattering (HLbL) contribution, shown in \Figref{hadronic_g_minus_2} (b),  is notoriously difficult to calculate. Nevertheless, phenomenological (data-driven, dispersive) and LQCD predictions have advanced to a precision better than $10\,\%$:
\begin{eqnarray}
a_\mu^\text{HLbL}(\text{phen.})&=&103.3(8.8)\times 10^{-11}, \label{phen_HLBL-WP25}\\
a_\mu^\text{HLbL}(\text{LQCD})&=&122.5(9.0)\times 10^{-11}.\label{LQCD_HLBL-WP25}
\end{eqnarray}
Therefore, the value of $a_\mu^\text{HLbL}$ recommended in WP25 \cite{Aliberti:2025beg},
\begin{subequations}
\label{WP25_HLBL}
\begin{equation}
a_\mu^\text{HLbL, LO}+a_\mu^\text{HLbL, NLO} = 115.5(9.9) \times 10^{-11} ,
\end{equation}
is built from the average of phenomenological and LQCD results at LO,
supplemented by the NLO contribution from phenomenology alone,
\begin{equation}
 a_\mu^\text{HLbL, NLO}=2.6(0.6)\times 10^{-11}.
 \end{equation}
 \end{subequations}
Due to improvements in the understanding of the subleading long-, mixed- and short-distance contributions from scalar, tensor, and axial-vector mesons, the HLbL contribution increased (staying within its uncertainty) relative to the WP20 estimate \cite{Aoyama:2020ynm}. At the same time, the leading contributions from pseudoscalar ($\pi, \eta, \eta'$) poles, ($\pi, K$) boxes, as well as $S$-wave $\pi\pi$ re-scattering, have undergone smaller shifts in the opposite direction.
 
\begin{figure}[t]
    \centering
    
\begin{tikzpicture}[baseline=0]
  \begin{feynman}
    \coordinate (mu1) at (-1.8, -1.2);
    \coordinate (v1)  at (-1.4, -1.2); 
    \coordinate (v2)  at (0.1, -1.2);  
    \coordinate (v3)  at (1.5, -1.2);  
    \coordinate (mu2) at (1.8, -1.2);
    
    \node[circle, fill=teal!15, draw=teal!80!black, thick, minimum size=0.5cm, inner sep=0pt] 
      (tff1) at (-1, 0.2) {};
      
    \node[circle, fill=teal!15, draw=teal!80!black, thick, minimum size=0.5cm, inner sep=0pt] 
      (tff2) at (0.8, -0.2) {};
      
    \coordinate (gamma) at (-1, 1.4);
    
    
    \draw[thick] (mu1) node[left] {\(\mu\)} -- (mu2) node[right] {\(\mu\)};
    
    \draw[photon] (tff1) -- (gamma) node[above] {\(\gamma\)};
    
    \draw[photon] (v1) -- (tff1);
    \draw[photon] (v2) -- (tff2);
    \draw[photon] (v3) -- (tff2);
    
    \draw[dashed, thick] (tff1) -- (tff2) node[midway, sloped, above=1pt] {\(\pi^0\)};
    
  \end{feynman}
\end{tikzpicture}
    \caption{Pion-pole contribution to $a_\mu$. The blobs denote the pion transition form factor. Crossed diagrams are not drawn.}
    \label{fig:pion_pole}
\end{figure}
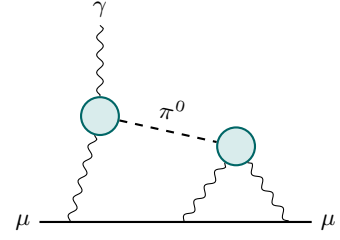

The largest HLbL contribution stems from the pion-pole diagram, $a_\mu^{\pi^0\text{-pole}} \sim (60\twodots 64) \times 10^{-11}$, shown in \Figref{pion_pole}. Its evaluation requires input for the interaction between a neutral pion and two photons, parametrized by the pion transition form factor (TFF), $\mathcal{F}_{\pi^0 \gamma^* \gamma^*} (Q_1^2, Q_2^2)$, with $Q_{1,2}^2=-q_{1,2}^2$ the virtualities of the two photons; the TFF is needed both in singly- and doubly-virtual photon kinematics. The electromagnetic decay of the neutral pion into two real photons,  $\pi^0 \rightarrow \gamma \gamma$, measured by  the PrimEx and PrimEx-II collaborations \cite{PrimEx:2010fvg,PrimEx-II:2020jwd}, provides a constraint for the normalization of the pion TFF. Scattering experiments have mapped out the pion TFF in the limit of one quasi-real and one virtual (space- or time-like) photon. A direct measurement of the pion TFF in doubly-virtual photon kinematics is prohibitively difficult; consequently, a data-driven description of the doubly-virtual pion TFF must rely on dispersion relations \cite{Hoferichter:2018kwz, Hoferichter:2018dmo}. For LQCD, the situation is quite the opposite: while LQCD can predict the pion TFF in general kinematics, it suffers from larger uncertainties at small photon virtualities. 

In this work, we perform a feasibility study for the combined analysis of LQCD and experimental data for the pion TFF. Combining constraints from complementary kinematics reduces uncertainties arising from extrapolations into unmeasured or poorly constrained regions. Furthermore, using physical world data as an anchor for the LQCD results reduces the uncertainty of the chiral and continuum extrapolations. Previous works that simultaneously included LQCD and experimental data in their analysis supplemented the experimental or the LQCD data with only a few selected data points from experiment or, respectively, LQCD. For instance, the Mainz/CLS Collaboration \cite{Gerardin:2019vio} included the measured pion decay width \cite{PrimEx:2010fvg} in their analysis. Conversely, the resonance chiral theory (RChT) fit of experimental data for the pseudoscalar TFFs by \citet{Estrada:2024cfy} included nine additional LQCD data points for the doubly-virtual case  ($Q_1^2=Q_2^2$ at $0.1, 1, 4$ GeV$^2$ for $\pi, \eta$, and $\eta'$, respectively) from the Budapest-Marseille-Wuppertal (BMW) Collaboration \cite{Gerardin:2023naa}. Here, we extend the analysis presented in Ref.~\cite{Gerardin:2019vio} by incorporating the existing experimental world data for the pion TFF in the space-like region. 

The paper is organized as follows. In \secref{GlobalConstraints}, we collect constraints on the pion TFF from LQCD, experiments, and theory. 
The fitting procedure, based on the modified $z$-expansion \cite{Gerardin:2019vio}, is outlined in \secref{Fittingprocedure}. In this section, a synthetic jackknife replicate sampling of the experimental data is introduced to allow for a statistically meaningful combination with the LQCD jackknife samples. Furthermore, we compare one- and two-stage fit approaches for the chiral and continuum extrapolations of the LQCD data. Results for $\mathcal{F}_{\pi^0 \gamma^* \gamma^*} (Q_1^2, Q_2^2)$ and $a_\mu^{\pi^0\text{-pole}}$ are presented in Sections \ref{sec:TFF} and \ref{sec:gm2}, respectively. We conclude with a summary and outlook in \secref{Summary}.

\section{Global constraints} \seclab{GlobalConstraints}

 The pion TFF $\mathcal{F}_{\pi^0 \gamma^* \gamma^*} (Q_1^2, Q_2^2)$ 
 is defined through the matrix element of two electromagnetic currents between the vacuum and the pion state:
\bea
&&\int d^4x \, e^{iq_1 \cdot x} \, \langle 0 | T j_\mu(x) j_\nu(0) | \pi^0(q_1 + q_2) \rangle \nn\\
&=& i e^2 \epsilon_{\mu\nu\alpha\beta} \, q_1^\alpha q_2^\beta \, \mathcal{F}_{\pi^0\gamma^*\gamma^*}(Q_1^2, Q_2^2),
\label{eq:TFF_definition}
\eea
where 
\begin{equation}
j_\mu = e \sum_{f} \mathcal{Q}_f \, \bar{q}_f \gamma_\mu q_f,
\label{eq:current}
\end{equation}
denotes the electromagnetic current carried by the quarks of  flavor $f$ and charge $\mathcal{Q}_f$. 

\begin{figure*}[t]
    \centering
 \includegraphics[width=1.0\textwidth]{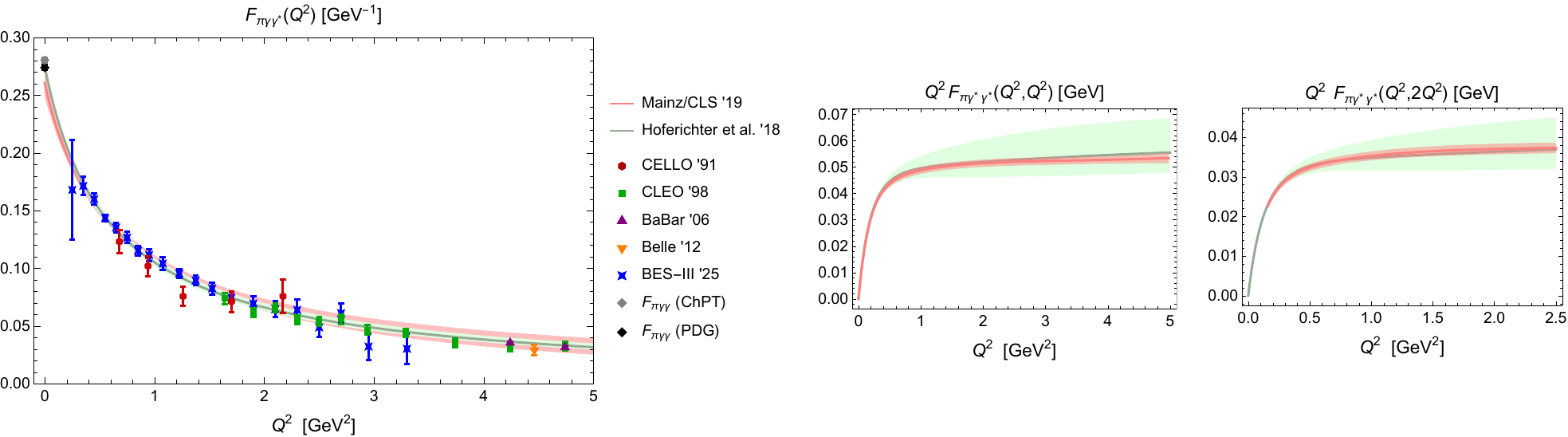}
    
    \caption{Comparison of selected results for the pion transition form factor in different kinematic regions: 
Mainz/CLS LQCD (pink band) \cite{Gerardin:2019vio}, dispersive representation (green band) \cite{Hoferichter:2018kwz,Hoferichter:2018dmo}, chiral perturbation theory (gray diamond) \cite{Kampf:2009tk}, normalization from PDG (black diamond) \cite{ParticleDataGroup:2024cfk}, and data from $e^+ e^-$ scattering \cite{Behrend:1990sr,Gronberg:1997fj,Aubert:2009mc,Uehara:2012ag,BESIII:2025zjx}.}
    \label{fig:IntroFigurePionTFF}
\end{figure*}

Following the WP20 standard, a description of the pion TFF should meet three criteria \cite{Aoyama:2020ynm}:
\begin{enumerate}
    \item in addition to the TFF normalization given by the real-photon decay widths, high-energy constraints must also be fulfilled;
\item at least the spacelike experimental data for the singly-virtual TFF must be reproduced;
\item systematic uncertainties must be assessed with a reasonable procedure.
\end{enumerate}
According to this standard, two independent data-driven approaches were considered in WP20: the dispersive prediction \cite{Hoferichter:2018kwz,Hoferichter:2018dmo}, and a Canterbury approximants fit \cite{Masjuan:2017tvw} of experimental data for the pseudoscalar-meson TFFs in the space-like region. Furthermore, an LQCD prediction of the pion TFF by the Mainz/CLS Collaboration \cite{Gerardin:2016cqj,Gerardin:2019vio} led to a result consistent with experimental data, as shown in \Figref{IntroFigurePionTFF}. This consistency is a necessary prerequisite for our combined analysis of experimental and LQCD data, which seeks to leverage both sources to improve the precision of $\mathcal{F}_{\pi^0 \gamma^* \gamma^*} (Q_1^2, Q_2^2)$ and $a_\mu^{\pi^0\text{-pole}}$. Our objective is to perform a joint analysis that adheres to the aforementioned WP20 criteria while incorporating the LQCD data from Ref.~\cite{Gerardin:2019vio} alongside the current experimental world data (including the recent BESIII results \cite{BESIII:2025zjx}) for the space-like TFF and the PrimEx-II decay width \cite{PrimEx-II:2020jwd}. In this section, we compile all global constraints on the pion TFF that serve as input for our analysis.

\subsection{Lattice QCD}

\begin{table}[tbp]
\centering
\caption{Parameters of the LQCD simulations \cite[Table 1]{Gerardin:2019vio}: the lattice resolution (periodic boundary conditions in $L$ and open boundary conditions in $T$), the bare coupling $\beta = 6/g_0^2$, the lattice spacing $a$ in physical units, the pion mass $m_\pi$, and the number of data points $N_\mathrm{data}$ contained in the $ll$ and $lc$ discretization ensembles, respectively.\label{tab:merged_sim_params}}
\begin{tabular*}{\columnwidth}{@{\extracolsep{\fill}}lllllll}
\hline
Id  & $L^3 \times T$ & $\beta$& $a$ [fm] & $m_\pi$ [MeV] & \multicolumn{2}{c}{$N_\mathrm{data}$}\\
 &  & &  & & $ll$ disc. &$lc$ disc.\\
\hline
H101    & $32^3 \times 96$  & 3.40& 0.08636 & 416(6) & 1617 & 1618\\
H102         & $32^3 \times 96$  &   &      & 354(5) & 3144 & 3148\\
N101         & $48^3 \times 128$ &    &     & 280(4) & 7370 & 7381\\
C101         & $48^3 \times 96$  &    &     & 224(3) & 3198 & 3201\\
\hline
S400    & $32^3 \times 128$ & 3.46& 0.07634 & 349(5) & 2808 & 2813\\
N401         & $48^3 \times 128$ &     &    & 286(4) & 6924 & 6936\\
\hline
N202     & $48^3 \times 128$&   3.55 &      0.06426    & 411(5) & 3013 & 3016 \\
N203         & $48^3 \times 128$ &     &    & 346(5) & 3130 & 3132\\
N200         & $48^3 \times 128$ &    &     & 284(3) & 3180 & 3183\\
D200         & $64^3 \times 128$ &     &    & 200(3) & 7385 & 7377\\
\hline
N300   & $48^3 \times 128$  & 3.70& 0.04981 & 422(5) & 2665 & 2665\\
J303         & $64^3 \times 192$ &    &     & 258(3) & 5115 & 5113\\
\hline
\end{tabular*}
\end{table}

The first LQCD prediction of the pion TFF was published by the Mainz/CLS Collaboration in 2016 \cite{Gerardin:2016cqj}. Since then, several other LQCD collaborations have reported results for the $\pi^0$, $\eta$ and $\eta'$ TFFs: the RBC-UKQCD Collaboration studied the $\pi^0$ TFF \cite{Lin:2024khg}, the Extended Twisted Mass (ETM) Collaboration evaluated  the $\pi^0$ and $\eta$ TFFs \cite{ExtendedTwistedMass:2023hin, ExtendedTwistedMass:2022ofm}, and the BMW Collaboration computed the $\pi^0$, $\eta$ and $\eta'$ TFFs \cite{Gerardin:2023naa}.

While our work can be extended, analogously to the empirical RChT fit in Ref.~\cite{Estrada:2024cfy}, into a global analysis of LQCD and experimental data for all light pseudoscalar-meson TFFs, we focus for now on the pion TFF. We utilize a subset of twelve $N_f=2+1$ Coordinated Lattice Simulations (CLS) \cite{Bruno:2014jqa}
ensembles with $\mathcal{O}(a)$-improved Wilson fermions and vector currents, using both local ($l$) and point-split ($c$) discretizations, and a tree-level $\mathcal{O}(a^2)$-improved Lüscher-Weisz action.  With the exception of the N302 ensemble, which is excluded here, this is the same dataset as in Ref.~\cite{Gerardin:2019vio}, which we refer the reader to for further details. The ensembles cover four different lattice spacings ($a\approx 0.050, 0.064, 0.076$, and $0.086$ fm) and
several pion masses down to $200$ MeV (D200). The scale setting has been performed with a precision of $1\,\%$ based on the pion and kaon decay constants \cite{Bruno:2016plf}. The parameters of the LQCD simulations are summarized in Table \ref{tab:merged_sim_params}.

The LQCD data cover space-like photon virtualities up to about $5.5$~GeV$^2$ (C101 and H101 reach $3.2$ and $4.2$~GeV$^2$, respectively) in the
doubly-virtual case, and up to $1.5$~GeV$^2$ in the singly-virtual limit \cite{Gerardin:2019vio}.\footnote{Data points with small time-like virtualities have been excluded from our analysis.} Compared to the first LQCD simulation \cite{Gerardin:2016cqj}, the kinematic reach in the $Q_1^2$ and $Q_2^2$ plane was improved by using 
both the pion rest and moving reference frames. While the pion rest frame reaches larger virtualities in the doubly-virtual kinematics, the moving reference frame does so in the singly-virtual kinematics. Furthermore, the pion rest frame yields statistically more precise data when both photons carry the same kinematics, whereas the moving reference frame is more precise in the singly-virtual kinematics. In general, the doubly-virtual kinematics are less sensitive to long-distance physics and, therefore, easier to calculate on the lattice. 

While an original, dedicated LQCD analysis extracts the pion TFF from simulated quark-connected and -disconnected contributions to the three-point correlation function, our work relies on the already extracted LQCD prediction for the pion TFF, $\mathcal{F}_{\pi^0 \gamma^* \gamma^*} (Q_1^2, Q_2^2)$, at different kinematic points. Regarding the error budget, statistical errors are estimated using the jackknife procedure, as discussed in \secref{jackknife}, while the uncertainty arising from neglected quark-disconnected diagrams is omitted here and only discussed in the context of $a_\mu$, see \secref{gm2}. Finite-size effects are expected to be small, since the TFF results from the pion rest and moving reference frames are consistent; nevertheless, ensembles H200 and H105 (the smaller counterparts of N202 and N101) were excluded in Ref.~\cite{Gerardin:2019vio} to be conservative. For a more detailed discussion of LQCD uncertainties, we refer the reader to Ref.~\cite{Gerardin:2019vio}.

\subsection{Scattering experiments}\seclab{scattering}
The singly-virtual pion TFF, $$\mathcal{F}_{\pi^0\gamma\gamma^*}(Q^2)=\mathcal{F}_{\pi^0\gamma^*\gamma^*}(Q^2,0)=\mathcal{F}_{\pi^0\gamma^*\gamma^*}(0,Q^2),$$ 
is measured by the CELLO \cite{Behrend:1990sr}, CLEO \cite{Gronberg:1997fj}, BaBar \cite{Aubert:2009mc}, Belle \cite{Uehara:2012ag}, and BESIII \cite{BESIII:2025zjx} collaborations in $e^+e^-\rightarrow e^+ e^- \pi^0$. The available world data, summarized in Table \ref{tab:TFFdatasets}, cover space-like photon virtualities from $0.25$ to $34.46$~GeV$^2$. Whenever possible, we add point-by-point statistical and systematic uncertainties in quadrature, and include the correlated $Q^2$-independent normalization (scaling) uncertainties, $\delta_\mathrm{norm}$, through scaling factors in our $\chi^2$ fits, as explained below. For Belle, the reported $6\,\%$ normalization uncertainty on the cross section corresponds to $3\,\%$ on the TFF; we subtract this value from their published systematic uncertainties \cite[Table 3]{Uehara:2012ag}.  BaBar published the combined statistical and systematic point-to-point uncertainties \cite[Table 2]{Aubert:2009mc}, and quotes an additional $Q^2$-independent normalization uncertainty of $2.3\,\%$. CELLO estimates a $12\,\%$ ($6\,\%$) systematic uncertainty on the cross section (TFF) \cite[Table 1]{Behrend:1990sr}, which we treat as a normalization uncertainty.  For CLEO, we observe a systematic uncertainty of $\sim 2.5\,\%$ across all points and treat it as a correlated $Q^2$-independent normalization uncertainty. For the most recent low-$Q^2$ data from BESIII \cite{BESIII:2025zjx}, information on the normalization uncertainty is not yet available; we use their smallest point-to-point systematic uncertainty of ($\sim 1.7\,\%$) as an estimate of the additional normalization uncertainty. 

\begin{figure*}[t]
    \centering
 \includegraphics[width=1.0 \textwidth]{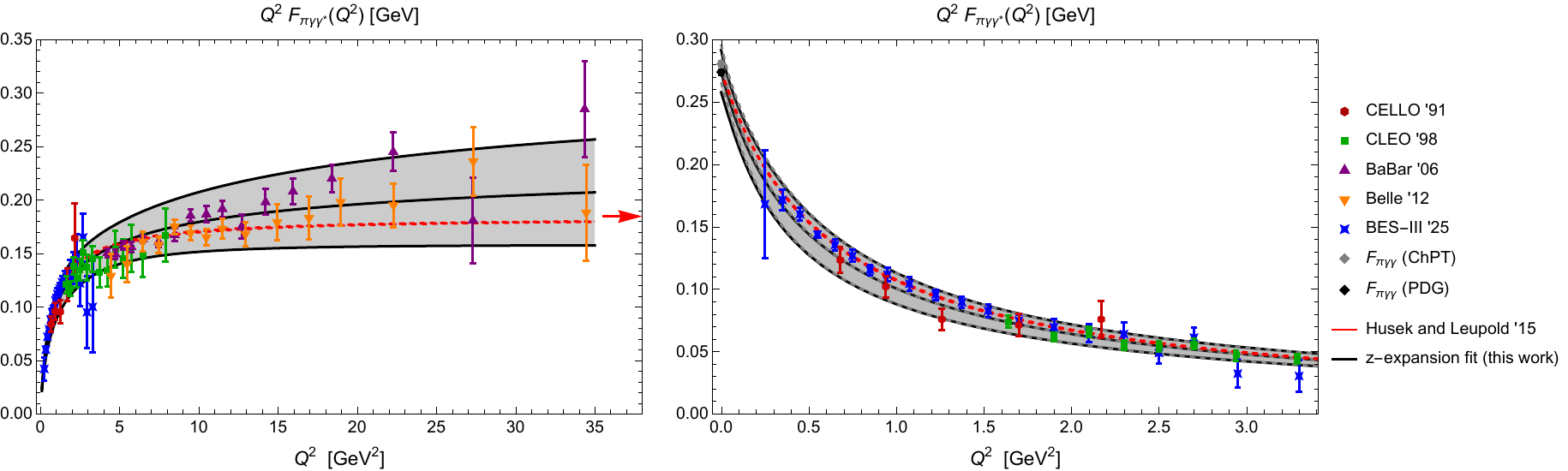}
    \caption{Comparison of our modified  $z$-expansion fits to the underlying experimental data. Details of the fits are presented in \secref{synthetic}. The two fits, utilizing either the theoretical (grey dashed) \eref{normalizationconstraintChPT} \cite{Kampf:2009tk} or the experimental (black solid) \eref{normalizationconstraintPDG} \cite{ParticleDataGroup:2024cfk} normalizations, differ only at very low $Q^2$. The red arrow indicates the Brodsky-Lepage limit \eref{BL}, and the red dashed band shows the prediction of the THS model \cite{Husek:2015wta}. }
    \label{fig:FitPlotPionTFF}
\end{figure*}

In Figures \ref{fig:IntroFigurePionTFF} and \ref{fig:FitPlotPionTFF}, we show the experimental data for the singly-virtual pion TFF compared to the LQCD prediction from \cite{Gerardin:2019vio}, the dispersive prediction \cite{Hoferichter:2018kwz,Hoferichter:2018dmo}, and the two-hadron saturation (THS) model \cite{Husek:2015wta}, which is predictive in the singly-virtual region as all its physical parameters are known. At large photon virtualities, a small tension exists between the data from BaBar and Belle. In \secref{synthetic}, we investigate the extent to which this tension is covered by the normalization uncertainties or if a systematic shift is observed when excluding BaBar from our fits.

In the future, analogous to the large-$N_c$ chiral perturbation theory (ChPT) fit in Ref.~\cite{Bickert:2020kbn}, our work might be extended to a simultaneous analysis of time- and space-like data. The time-like pion TFF is measured through the $\pi^0\rightarrow e^+e^-\gamma$ Dalitz decay by the A2 \cite{A2:2016sjm,A2:2025srw} and NA62 \cite{NA62:2016zfg} collaborations, as well as in $e^+e^-\rightarrow \pi^0 \ga$ by the CMD-2 \cite{CMD-2:2004ahv} and SND  \cite{SND:2016drm} collaborations.

\begin{table}[t]
\centering
\caption{Experiments measuring the space-like singly-virtual pion transition form factor:
$\mathcal{F}_{\pi^0\gamma\gamma^*}(Q^2)$. Values
with a single asterisk $(^*)$ are educated guesses for the normalization uncertainties assigned by the authors of this work (see text for details).\label{tab:TFFdatasets}}
\renewcommand{\arraystretch}{1.2}
\begin{tabular*}{\columnwidth}{@{\extracolsep{\fill}}l c c c}
\hline
Experiment & $Q^2$ [GeV$^2$] &  $N_{\mathrm{data}}$ & $\delta_{\mathrm{norm}}\,[\%]$ \\
\hline
CELLO \cite{Behrend:1990sr}  & 0.68 -- 2.17  &  5 & 6 \\
CLEO \cite{Gronberg:1997fj}   & 1.64 -- 7.90  & 15 & 2.5$^{*}$ \\
BaBar \cite{Aubert:2009mc}   & 4.24 -- 34.36 & 17 & 2.3\\
Belle \cite{Uehara:2012ag}   & 4.46 -- 34.46 & 15 & 3 \\
BESIII \cite{BESIII:2025zjx} & 0.25 -- 3.3   & 20 & 1.7$^{*}$ \\
\hline
\end{tabular*}
\renewcommand{\arraystretch}{1}
\end{table}

\subsection{Pion decay} \seclab{piondecay}

The dominant decay channel of the neutral pion is the electromagnetic decay into two real photons,  $\pi^0 \rightarrow \gamma \gamma$, with a branching ratio of
\beq
\mathrm{Br}(\pi^0 \to \gamma\gamma)=98.823(34)\,\%.
\eeq
The associated decay width has been measured by the 
PrimEx \cite{PrimEx:2010fvg} and PrimEx-II  \cite{PrimEx-II:2020jwd} collaborations at Jefferson Laboratory (JLab) with a combined accuracy of $1.5\,\%$, see Table \ref{tab:Decaydata}. The neutral pion lifetime recommended by the Particle Data Group (PDG) \cite{ParticleDataGroup:2024cfk}:
\beq
\tau=8.43(13)\times 10^{-17}\mathrm{s},
\eeq
is an average of five experiments: two Primakoff measurements (one at Cornell University \cite{Browman:1974cu} and aforementioned measurements at JLab \cite{PrimEx:2010fvg,PrimEx-II:2020jwd}), a direct measurement
from CERN \cite{Atherton:1985av}, a collider measurement \cite{CrystalBall:1988xvy},
and a derivation from the radiative pion beta
decay \cite{Bychkov:2008ws}. The two most precise determinations from CERN and JLab differ by $6\,\%$ \cite{PrimEx-II:2020jwd}.

The electromagnetic decay width,
\begin{equation}
   \Gamma(\pi^0 \to \gamma\gamma) = \frac{M_\pi^3e^4}{64 \pi} 
   \vert\mathcal{F}_{\pi^0\gamma\gamma}\vert^2,
   \eqlab{decaywidth}
\end{equation}
is proportional to the square of the pion TFF in the real-photon limit: 
\begin{equation}
\mathcal{F}_{\pi^0\gamma^*\gamma^*}(0,0) =\mathcal{F}_{\pi^0\gamma\gamma}.
\end{equation}
In the chiral limit, the low-energy theorem completely dictates the normalization of the TFF via the chiral anomaly \cite{Adler:1969gk,Bell:1969ts,Bardeen:1969md}:
\begin{equation}
\mathcal{F}_{\pi^0\gamma\gamma} \sim \frac{1}{4\pi^2 F},
\end{equation}
where $F$ is the pion decay constant in the chiral limit. We denote the value of the pion decay constant at unphysical pion masses by $f_\pi$. The physical pion decay constant, $F_\pi$, is determined through the charged pion decay \cite{ParticleDataGroup:2024cfk}: 
\beq
F_\pi=92.320(97)\,\mathrm{MeV}. \eqlab{Fpi}
\eeq

While theoretical predictions for the normalization of the pion TFF from ChPT are available with \mbox{(sub-)percent-level} precision \cite{Goity:2002nn,Ananthanarayan:2002kj,Kampf:2009tk}, they differ from the currently most precise experimental constraint from PrimEx-II \cite{PrimEx-II:2020jwd} by more than $2\,\sigma$, as detailed in Table \ref{tab:Decaydata}. In \secref{TFF}, we investigate the systematic uncertainty arising from this discrepancy and its ultimate impact on the muon $g-2$ through separate fits incorporating either the experimental or the theoretical value as a constraint on the normalization:
\begin{subequations}
\eqlab{normalizationconstraint}
    \bea
    \mathcal{F}_{\pi^0\gamma\gamma}(\mathrm{PDG})&=& 0.2739(21)\,\mathrm{GeV}^{-1}\, \text{\cite{ParticleDataGroup:2024cfk},}\eqlab{normalizationconstraintPDG}\\
        \mathcal{F}_{\pi^0\gamma\gamma}(\text{ChPT NNLO})&=& 0.2805(19)\,\mathrm{GeV}^{-1} \, \text{\cite{Kampf:2009tk}.} \eqlab{normalizationconstraintChPT}\qquad
    \eea
\end{subequations}

\begin{table}[t]
\centering
\caption{Experimental values and theoretical predictions for the decay width of the neutral pion into two real photons, $\Gamma(\pi^{0} \to \gamma\gamma)$, and the corresponding values for the pion transition form factor in the real-photon limit following \Eqref{decaywidth}.\label{tab:Decaydata}}
\renewcommand{\arraystretch}{1.2}
\begin{tabularx}{\columnwidth}{XXX}
\hline
 & $\Gamma(\pi^{0} \to \gamma\gamma)$ [eV] & $\mathcal{F}_{\pi^0\gamma\gamma}$ [GeV$^{-1}$]\\
  \hline
 \textit{Experiment} & & \\
 PDG & 7.72(12) & 0.2739(21)\\
 PrimEx-II \cite{PrimEx:2010fvg,PrimEx-II:2020jwd} & 7.802(117) & 0.2754(21) \\
 CERN & 7.25(23) & 0.266(4) \\
 \hline
 \textit{ChPT} & & \\
 GBH (NLO) \cite{Goity:2002nn} & 8.10(8) & 0.2806(14)\\
 AM (NLO) \cite{Ananthanarayan:2002kj} & 8.06(6) & 0.2799(10)\\
 KM (NNLO) \cite{Kampf:2009tk} & 8.09(11) & 0.2805(19)\\
  \hline
 \textit{QCD sum rules} & & \\
 IO \cite{Ioffe:2007eg}& 7.93(12) & 0.278(2)\\
\hline
 \textit{Chiral anomaly} & &\\
  $F_\pi$ from PDG \eref{Fpi}& 7.743(16)& 0.2744(3)\\
  \hline
\end{tabularx}
\renewcommand{\arraystretch}{1}
\end{table}

\subsection{Asymptotic constraints} \seclab{asymptotic}

Further constraints on the pion TFF can be derived in the limit of asymptotically large photon virtualities. We consider the well-known  
Brodsky-Lepage (BL) limit of the singly-virtual pion TFF \cite{Lepage:1980fj,Brodsky:1981rp} and a constraint on the symmetric pion TFF from the operator product expansion (OPE) \cite{Nesterenko:1982dn,Novikov:1983jt}:
\begin{subequations}
\eqlab{asymptoticsEq}
    \bea
    \lim_{Q^2\to\infty}  \mathcal{F}_{\pi^0\gamma\gamma^*}(Q^2) &=& \frac{2F_\pi}{Q^2},\eqlab{BL}\\
    \lim_{Q^2\to\infty}\mathcal{F}_{\pi^0\gamma^*\gamma^*}(Q^2, Q^2) &=& \frac{2F_\pi}{3 Q^2 }\left(1-\frac{8}{9}\frac{\delta_\pi^2}{Q^2}\right),\eqlab{OPEsym}\qquad
\eea
\end{subequations}
where the higher-order correction $\delta_\pi^2=0.20(2)\,\text{GeV}^2$ has been determined from QCD sum rules
 \cite{Novikov:1983jt}. We implement these constraints on the physical pion TFF as additional points in the $\chi^2$, see \secref{penalty}.

\section{Fitting procedure} \seclab{Fittingprocedure}

In this section, we outline the fitting procedure for the combined analysis of LQCD and experimental data. 

\subsection{Modified $z$-expansion} \seclab{zexpansion}
To parametrize the pion TFF and fit the global constraints collected in \secref{GlobalConstraints}, we employ the modified $z$-expansion proposed in \cite{Gerardin:2019vio}:\footnote{To estimate model dependence, results were compared to fits based on the simple LMD+V model and Canterbury approximants.}
\begin{equation}
\eqlab{zexp}
\begin{split}
    &P(Q_1^2, Q_2^2) \mathcal{F}_{\pi^0\gamma^*\gamma^*}(Q_1^2, Q_2^2) \\
    &= \sum_{n,m=0}^N c_{nm} \bigg[ z_1^n - (-1)^{N+n+1} \frac{n}{N+1} z_1^{N+1} \bigg] \\
    &\quad \times \bigg[ z_2^m - (-1)^{N+m+1} \frac{m}{N+1} z_2^{N+1} \bigg],
\end{split}
\end{equation}
with the conformal variables $z_k$ ($k = 1, 2$) defined through
\begin{subequations}
\eqlab{zdef}
\bea 
    z_k &=& \frac{\sqrt{t_c + Q_k^2} - \sqrt{t_c - t_0}}{\sqrt{t_c + Q_k^2} + \sqrt{t_c - t_0}},\\
    t_0 &=& t_c \left( 1 - \sqrt{1 + Q_{\max}^2 / t_c} \right).
\eea 
\end{subequations}
The $z$-expansion maps the branch cut starting at $t_c$ onto the unit circle $\vert z_k\vert=1$. In the physical world, this branch cut is generated by a charged pion pair. However, since the pion masses are degenerate in the isospin-symmetric lattice simulations, we choose 
\beq
  t_c=4M_{\pi,\text{iso}}^2,
\eeq
with $M_{\pi,\text{iso}}^2$ close to the neutral pion mass \cite{Djukanovic:2023beb}:
\beq
M_{\pi,\text{ phys}}=M_{\pi,\text{ iso}}=134.8(3)\, \text{MeV}. \eqlab{mpiiso}
\eeq
The analytical function
\begin{equation}
P(Q_1^2, Q_2^2)= 1 + \frac{Q_1^2 + Q_2^2}{M_\rho^2}\,,\eqlab{Pdef}
\end{equation}
 where $M_\rho=775.26(23)\,\mathrm{MeV}$ is the physical $\rho$ meson mass, ensures that the modified $z$-expansion decreases asymptotically as $1/Q^2$ in all directions in the $(Q_1^2,Q_2^2)$ plane, in accordance with the BL and OPE limits defined in Eqs.~\eref{BL} and \eref{OPEsym}.

The modified $z$-expansion offers a flexible ansatz well suited to validate the methodology of our combined analysis. 
Furthermore, it allows us to directly compare our results with the original analysis of the employed LQCD dataset \cite{Gerardin:2019vio}. The authors of Ref.\ \cite{Gerardin:2019vio} concluded that setting $N=3$ and $Q_\mathrm{max}^2=4\,\mathrm{GeV}^2$ was sufficient to describe the LQCD data on the pion TFF 
with sub-percent precision. In \secref{synthetic}, we study the truncation of the expansion in \eref{zexp} and the dependence on $Q_\mathrm{max}^2$ for the experimental data, which cover a wider $Q^2$ range. 

\subsection{Jackknife procedure} \seclab{jackknife}

The LQCD data are provided in the form of jackknife replicates. We use the jackknife procedure to estimate the statistical error of the pion TFF and, subsequently, of the fit parameters. Consider the pion TFF at a specific kinematic point, $\mathcal{F}_{\pi^0 \gamma^* \gamma^*} (Q_1^2, Q_2^2)$, denoted in the following by the shorthand $\mathcal{F}$. Its central value and 
 uncertainty, $\overline{\mathcal{F}} \pm \Delta \mathcal{F}$ , are defined as
\begin{subequations}
\eqlab{JKTFF}
\bea
\overline{\mathcal{F}} &=& \frac{1}{N_\mathrm{Jk}} \sum_{i=1}^{N_\mathrm{Jk}} \mathcal{F}^{(i)},\eqlab{mean}\\
\Delta \mathcal{F} &=& \frac{N_\mathrm{Jk}-1}{N_\mathrm{Jk}} \sum_{i=1}^{N_\mathrm{Jk}}  
\left[\mathcal{F}^{(i)}- \overline{\mathcal{F}}\right]^2,
\eea
\end{subequations}
where $N_\mathrm{Jk}=50$ is the number of jackknife replicates $\mathcal{F}^{(i)}$. We perform $N_\mathrm{Jk}$ independent fits, one for each set of jackknife replicates.\footnote{The correlation matrices of the LQCD data are ill-conditioned due to the large size of the datasets. Following Ref.~\cite{Gerardin:2019vio}, we therefore perform uncorrelated fits.}  Similar to \Eqref{JKTFF}, for a fit ansatz with $N_\mathrm{par}$ parameters $p_j$, we define their central values and covariance matrix as:
\begin{subequations}
\eqlab{JKpar}
\bea
\overline p_j &=& \frac{1}{N_\mathrm{Jk}} \sum_{i=1}^{N_\mathrm{Jk}} p_j^{(i)},\eqlab{mean_jackknife}\\
\Sigma_{jl} &=& \frac{N_\mathrm{Jk}-1}{N_\mathrm{Jk}} \sum_{i=1}^{N_\mathrm{Jk}} (p_j^{(i)} - \bar{p}_j)(p_l^{(i)} - \bar{p}_l),
\eea
\end{subequations}
where $p_j^{(i)}$ is the $j$-th parameter fitted to the $i$-th set of jackknife replicates. The parameter covariance matrix $\Sigma_{jl}$ is a symmetric matrix of dimension $N_\mathrm{par}\times N_\mathrm{par}$.
The standard errors are obtained from the square roots of the diagonal elements of $\boldsymbol{\Sigma}$, while the off-diagonal elements capture the correlations propagated through the fitting procedure. This approach ensures that the statistical relationships between parameters are preserved and that the uncertainties correctly reflect the variance of the underlying data.

\subsection{Synthetic data jackknife replicate sampling} \seclab{synthetic}

It is important to combine LQCD and experimental data in a statistically meaningful way. As explained in the previous section, the statistical errors of the LQCD data are estimated using $50$ jackknife replicates. To perform a simultaneous fit of the LQCD and experimental data, we must generate $50$ corresponding experimental samples. Since the correlations between measurements at different virtualities and from different experiments are unknown, we perform a synthetic data jackknife replicate sampling of the experimental world data to avoid artificial fluctuations between (neighboring) points. Unlike a standard non-parametric sampling of the experimental data, the synthetic approach samples new data points not directly from the original data, but from a fit to the original data, varying the fit parameters within their uncertainties. 

To fit the experimental data, we minimize the following $\chi^2$ function:
\bea
\chi_{\text{exp.}}^2&=&\sum_{j=1}^{N_\text{exp.}} \left\{\sum_{i=1}^{N_{\text{data}, j}} \left[\frac{k_j\,\mathcal{F}^\text{(exp.)}_{\pi^0\ga\ga^*}(Q_i^2)-\mathcal{F}^\text{(th.)}_{\pi^0\ga\ga^*}(Q_i^2)}{k_j\,\Delta \mathcal{F}^\text{(exp.)}_{\pi^0\ga\ga^*}(Q_i^2)}\right]^2\right.\nn\\
&&\left.+\left(\frac{k_j-1}{k_j \delta_{\text{norm}, j}}\right)^2\right\}+\left[\frac{\mathcal{F}^\text{(exp.)}_{\pi^0\ga\ga}-\mathcal{F}^\text{(th.)}_{\pi^0\ga\ga}}{\Delta \mathcal{F}^\text{(exp.)}_{\pi^0\ga\ga}}\right]^2.\eqlab{chi2exp}
\eea 
This function represents the sum over $N_\text{exp.}=5$ scattering experiments, detailed in \secref{scattering} and Table \ref{tab:TFFdatasets}, alongside a constraint on the normalization from the pion decay width, $\Gamma(\pi^{0} \to \gamma\gamma)$, detailed in \secref{piondecay} and Table \ref{tab:Decaydata}.  Here, $\mathcal{F}^\text{(exp.)}_{\pi^0\ga\ga^{(*)}}$ and $\Delta \mathcal{F}^\text{(exp.)}_{\pi^0\ga\ga^{(*)}}$ denote the experimental points and their uncertainties, while $\mathcal{F}^\text{(th.)}_{\pi^0\ga\ga^{(*)}}$ is the theoretical model being adjusted. For each scattering experiment $j$, an overall scaling factor $k_j$ is fixed during the minimization to take into account the normalization uncertainty $\delta_{\text{norm}, j}$ of its $N_{\text{data}, j}$ data points. No penalty terms associated with asymptotic constraints on the pion TFF, detailed in \secref{asymptotic}, are included in this fit.

The suitability of a theoretical fit model depends not only on the observable but also on the available data. This becomes evident when examining the modified $z$-expansion and comparing the LQCD and experimental fits. As explained in \secref{zexpansion}, the LQCD data are well described by \Eqref{zexp} with $N=3$ and $Q_\mathrm{max}^2=4\,\mathrm{GeV}^2$ \cite{Gerardin:2019vio}, yielding ten $c_{nm}$ fit parameters. While the LQCD data cover the doubly-virtual region for $Q^2\leq 5.5$~GeV$^2$, the $72$ experimental points from $e^+e^-$ scattering cover much larger photon virtualities, $Q^2\in [0.25,34.46]$~GeV$^2$, but only in the singly-virtual photon limit. In this limit, the $c_{nm}$ parameters become linearly dependent, and only $N+1$ independent parameters remain. It is therefore important to evaluate which choice of $N$ and $Q_\mathrm{max}^2$ is suitable to describe the experimental data in both a standalone and a combined fit. 

The choice of $Q_\mathrm{max}^2$ limits the size of $\vert z_k\vert$, defined in \Eqref{zdef}, within a given $Q^2$ range. For $Q_\mathrm{max}^2=4\,\mathrm{GeV}^2$, we find $\vert z_k\vert<0.52$ in the region of the LQCD data, and $\vert z_k\vert<0.78$ in the region of the experimental data. A standalone fit of the experimental data showed that the modified $z$-expansion is already converging for $Q_\mathrm{max}^2=4\,\mathrm{GeV}^2$, thus, no increase of the $Q_\mathrm{max}^2$ parameter is required to incorporate the experimental data in a combined fit.

We study different subsets of the experimental data for three choices of $N=1,2,3$ in the modified $z$-expansion. At large $Q^2$, data from BaBar and Belle exhibit a small tension. Although systematically including and excluding these sets shows a slight preference for the Belle data, the BaBar measurements remain compatible within $\sim 3\,\sigma$ in a global fit, in which all scaling factors remain within their expected bounds dictated by the normalization uncertainties. Therefore, all available world data on the space-like pion TFF  from scattering experiments are included in our analysis. Furthermore, we study the systematic effect of constraining the TFF normalization using either the measured pion decay width or its ChPT prediction, see \Eqref{normalizationconstraint}. Both normalizations are compatible with the scattering data. To this end, we prepare two separate sets of experimental samples, whose influence on the combined analysis of LQCD and experimental data will be examined in subsequent sections. Our main results rely on the PDG recommended value \eref{normalizationconstraintPDG}. Remarkably, even a two-parameter $N=1$ ansatz describes the world data with a $\chi^2_\mathrm{exp}/N_\mathrm{dof}=1.65$, while the $N=2$ and $N=3$ fits yield $\chi^2_\mathrm{exp}/N_\mathrm{dof}=1.10$ and $0.90$, respectively. Since the $N=3$ fit is overfitting the data, we base our synthetic data sampling on the $N=2$ and $Q_\mathrm{max}^2=4\,\mathrm{GeV}^2$ fits shown in \Figref{FitPlotPionTFF}. For comparison, we display the THS model \cite{Husek:2015wta} as a prediction of the singly-virtual pion TFF in the low-$Q^2$ region.

Finally, we generate $N_\mathrm{Jk}=50$ synthetic jackknife replicates of the experimental world data, as illustrated in Figs.~\ref{fig:ExperimentalJKsamplesNormPlot}-\ref{fig:ExperimentalJKsamplesNormalization_3} in Appendix \ref{sec:supplement}. The three independent parameters of the ($N=2$) $z$-expansion are drawn from a multivariate normal distribution centered on their best-fit values. To ensure that the standard jackknife variance estimator correctly reproduces the statistical uncertainties of the original fit, the covariance matrix of the best-fit is scaled by a factor of $(N_\mathrm{Jk}-1)^{-1}$ to generate the appropriate multivariate distribution. For each replicate, the synthetic dataset is obtained by evaluating the $z$-expansion with the sampled parameters at the exact kinematic points of the original experiments. Consequently, the mean of the generated samples for each data point carries an uncertainty, evaluated via \Eqref{JKTFF}, that matches the uncertainty derived from the original fit using standard error propagation.

\subsection{Chiral and continuum extrapolation} \seclab{extrapolation}

Analogously to the LQCD analysis in Ref.~\cite{Gerardin:2019vio}, we base the combined analysis of LQCD and experimental data on a modified $z$-expansion with $N=3$ and $Q_\mathrm{max}^2=4\,\mathrm{GeV}^2$. For each LQCD ensemble, the pion TFF can be described through a coefficient set $c_{nm}(a, \tilde{y})$, where $a$ is the lattice spacing and $\tilde{y} = m_\pi^2 / (16\pi^2 F_\pi^2)$ is a dimensionless parameter proportional to the squared pion mass on the lattice. At the physical point, this parameter takes the value $\tilde{y}_{\text{phys}}=0.01350(7)$. The chiral and continuum extrapolation of the parameters is given by the functional form \cite{Gerardin:2019vio}:
\begin{equation}
\eqlab{Eqextrapol}
    c_{nm}(a, \tilde{y})= c_{nm} + \gamma_{nm}(\tilde{y} - \tilde{y}_{\text{phys}}) + \delta_{nm}^{(d)} \left( \frac{a}{a_{\beta=3.55}} \right)^2,
\end{equation}
where the coefficient for the continuum extrapolation, $\delta_{nm}^{(d)}$, differs between the two discretizations $d$. The experimental world data are then described by the physical parameters $c_{nm}$. For $N=3$, the modified $z$-expansion has ten coefficients. Including the chiral and continuum extrapolations, and without imposing asymptotic constraints from perturbative QCD, this number increases to $40$ free parameters for the combined fit of all LQCD ensembles (incorporating both $ll$ and $lc$ discretizations, see Table \ref{tab:merged_sim_params}).

\subsection{Additional penalty terms}\seclab{penalty}

In accordance with the WP20 quality criteria, quoted at the beginning of \secref{GlobalConstraints}, a parametrization of the pion TFF should fulfill the high-energy constraints discussed in \secref{asymptotic}. While phenomenological models and effective-field-theory predictions --- such as the THS model \cite{Husek:2015wta} or RChT ans\"atze \cite{Estrada:2024cfy,Kadavy:2022scu} --- explicitly comply with the BL and OPE limits introduced in \Eqref{asymptoticsEq},\footnote{See the review~\cite{Danilkin:2019mhd} for further discussions.} the modified $z$-expansion does not. Although it yields the correct $1/Q^2$ scaling for asymptotically large values of the photon virtualities $Q_1^2$ and $Q_2^2$, it does not fix the exact asymptotic coefficients. Furthermore, the large-$Q$ limit of the modified $z$-expansion incorrectly features odd powers of $1/Q$. 

To address these shortcomings, \secref{TFF} compares two distinct sets of fits: an unconstrained fit (following Ref.~\cite{Gerardin:2019vio}), as well as a constrained fit in which the asymptotic constraints from \secref{asymptotic} are imposed through additional penalty terms in the $\chi^2$ function. Additionally, we eliminate the spurious $1/Q^3$ contribution in the OPE limit. This constraint leads to an additional linear relation between $c_{nm}(0,\tilde y_\mathrm{phys})$, and we chose to express $c_{33}(0,\tilde y_\mathrm{phys})$ via the other parameters at the same point as
\beq
c_{33}=-(4 c_{10} + 5 c_{11} + 2 c_{21} +12  c_{30} +22 c_{31} +6 c_{32})/21\,.\eqlab{c33}
\eeq

\subsection{One- and two-stage fitting approaches}

In the context of chiral and continuum extrapolations of LQCD data \cite{Gerardin:2016cqj}, it is customary to evaluate and compare both one- and two-stage fitting approaches. In a preliminary feasibility study based on the LMD+V model \cite{Knecht:2001qf} for the pion TFF, we explored four distinct fitting methodologies for the combined analysis of the experimental world data and the $24$ LQCD ensembles ($12$ ensembles each for the $ll$ and $lc$ discretizations, see Table \ref{tab:merged_sim_params}). These correspond to four combinations of the two extrapolation strategies (one-stage vs.\ two-stage) and different modifications of the $\chi^2$ functions (introducing correlations and a normalized dataset reweighting). In the two-stage approach, the theoretical ansatz is first fitted individually to each of the $24$ LQCD ensembles and to the experimental world data. For each model parameter, this yields $25$ independent best-fit determinations at distinct values of $a$ and $\tilde{y}$, cf.\ the left-hand side of \Eqref{Eqextrapol}. Subsequently, a simultaneous chiral and continuum extrapolation of all parameters, taking into account their correlated uncertainties from the individual ensemble fits, proved significantly more robust than performing separate extrapolation fits for individual parameters. In fact, performing such a separate two-stage fit is fundamentally impossible for the modified $z$-expansion, as its parameters become linearly dependent in the singly-virtual limit covered by the experimental data. We therefore chose the global one-stage approach with a normalized $\chi^2$ weighting scheme, detailed in \Eqref{chi2weighting}, as our primary method. In this approach, we perform a single global fit of all LQCD ensembles and the experimental data to determine the physical $z$-expansion coefficients $c_{nm}(0,\tilde y_\mathrm{phys})$ alongside the extrapolation parameters $\gamma_{nm}$, $\delta_{nm}^{(ll)}$, and $\delta_{nm}^{(lc)}$. While fitting only singly-virtual experimental data reduces the number of linearly independent coefficients in the $z$-expansion, a simultaneous combined fit with doubly-virtual LQCD data preserves the independence of all coefficients. Thus, the singly-virtual data serve to further restrict the overall parameter space by tightly constraining parameter correlations.  

The fundamental building block for our fits is the $\chi^2$ function of an individual dataset $e$ (such as a LQCD ensemble) with $N_{e}$ data points, defined as
\beq
\chi_e^2=\sum_{i=1}^{N_e} \left\{\frac{\left[\mathcal{F}_{\pi^0\ga^*\ga^*}-\mathcal{F}^\text{(th.)}_{\pi^0\ga^*\ga^*}\right]\!(Q_{1,i}^2,Q_{2,i}^2)}{\Delta \mathcal{F}_{\pi^0\ga^*\ga^*}(Q_{1,i}^2,Q_{2,i}^2)}\right\}^2, \eqlab{chi2lattice}
\eeq
where  $\mathcal{F}$ represents either the jackknife mean or a jackknife replicate at a given kinematic point $(Q_{1,i}^2,Q_{2,i}^2)$. Following \secref{jackknife}, we evaluate the parameter covariance matrix   based on $N_\text{Jk}=50$ fits to each set of jackknife replicates. While the two-stage approach permits a direct minimization of the experimental world data via \Eqref{chi2exp} without resampling, the one-stage approach requires the simultaneous treatment of all datasets. Therefore, the synthetic jackknife replicate samples of the experimental world data, generated in \secref{synthetic}, are exclusively utilized in the one-stage fit, where they are utilized analogously to \Eqref{chi2lattice}.

For the global one-stage fit, the individual $\chi_e^2$ contributions from the LQCD ensembles and the resampled experimental data must be appropriately balanced. To achieve this, we employ a weighted $\chi^2$ ansatz, inspired by the approach utilized in Ref.~\cite{Pittler:2025upn} for the global analysis of LQCD and experimental data on negative-strangeness meson-baryon scattering. Our modified $\chi^2$ normalizes the contribution from each dataset, thereby preventing the LQCD ensembles --- which together contain three orders of magnitude more data points than the experiments --- from artificially dominating the fit. The reduced $\chi^2$ function is defined as:
\begin{equation}
\eqlab{chi2weighting}
\chi^2_{\text{dof}}=\frac{\sum_{e=1}^E N_e}{E \left[ \left( \sum_{e=1}^E N_e \right) -N_{\text{par}}\right]}\sum_{e=1}^E \frac{\chi^2_e}{N_e},
\end{equation} 
where $E=25$ is the number of individual datasets and $N_\text{par}$ is the total number of free fit parameters, including extrapolation parameters. To further increase the weight of the experimental data, one could combine the $ll$ and $lc$ discretizations, resulting in $E=13$ independent sets ($12$ LQCD sets plus $1$ experimental set). If penalty terms are imposed, we include them as a separate dataset.

\section{Combined analysis} \seclab{TFF}

To establish a baseline for our combined analysis of LQCD and experimental data, we first perform a standalone fit ({\bf Fit A}) of the LQCD dataset. While our findings are consistent with the results in Ref.~\cite{Gerardin:2019vio}, we obtain slightly larger uncertainties, particularly at low $Q^2$. As an illustration, we compare the extracted normalizations:
\begin{subequations}
\eqlab{normcomp}
\bea
\mathcal{F}_{\pi^0\gamma\gamma}(\text{Mainz/CLS})&=&0.261(13)_{\text{stat}}(2)_{\text{syst}}\,\mathrm{GeV}^{-1},\qquad \\
\mathcal{F}_{\pi^0\gamma\gamma}(\text{\bf Fit A})&=&0.254(21)_{\text{stat}}\,\mathrm{GeV}^{-1}.
\eea
\end{subequations}
Here, the first uncertainty is statistical, while the second estimates the systematic error associated with truncating the $z$-expansion. Note that while the Mainz/CLS normalization is fully compatible with both the PDG and the ChPT constraints, \Eqref{normalizationconstraint}, our fit exhibits a mild $1\,\sigma$ tension with the latter. 

Several factors contribute to our enlarged uncertainties. First, their dataset includes one additional fine ensemble: N302 with $a\sim 0.050$\,fm and $m_\pi =343(5)$\,MeV. Furthermore, we consider the entire $Q^2$ dataset without imposing kinematic cuts that are used to control systematic and statistical uncertainties (e.g., in the extraction of the slope parameter $b_\pi$). This difference, however, does not impede our primary objective, which is to quantify the relative improvement that can be achieved through a combined analysis of LQCD and experimental data.

To assess systematic effects, we perform four combined LQCD and experimental fits. All fits utilize the modified $z$-expansion with $N=3$ and $Q_\mathrm{max}^2=4\,\mathrm{GeV}^2$. They differ in the applied normalization constraint and the penalty terms that enforce the BL and OPE limits from \Eqref{asymptoticsEq}:
\begin{itemize}[leftmargin=1.25cm, labelwidth=3cm, align=right]
\item[{\bf Fit B1}:] PDG normalization \eref{normalizationconstraintPDG} and no penalty terms; 
\item[{\bf Fit B2}:] ChPT normalization \eref{normalizationconstraintChPT} and no penalty terms;
\item[{\bf Fit C1}:] PDG normalization \eref{normalizationconstraintPDG} and penalty terms; 
\item[{\bf Fit C2}:] ChPT normalization \eref{normalizationconstraintChPT} and penalty terms.
\end{itemize}
Including the extrapolation parameters, the fit models generally contain $40$ free parameters. However, {\bf Fits C1} and {\bf C2} eliminate $c_{33}$ following \Eqref{c33}, and therefore only has $39$ parameters. The best-fit parameters and correlation matrices are provided in Appendix \ref{sec:supplement}.

\begin{figure*}[t]
    \centering
 \includegraphics[width=0.9 \textwidth]{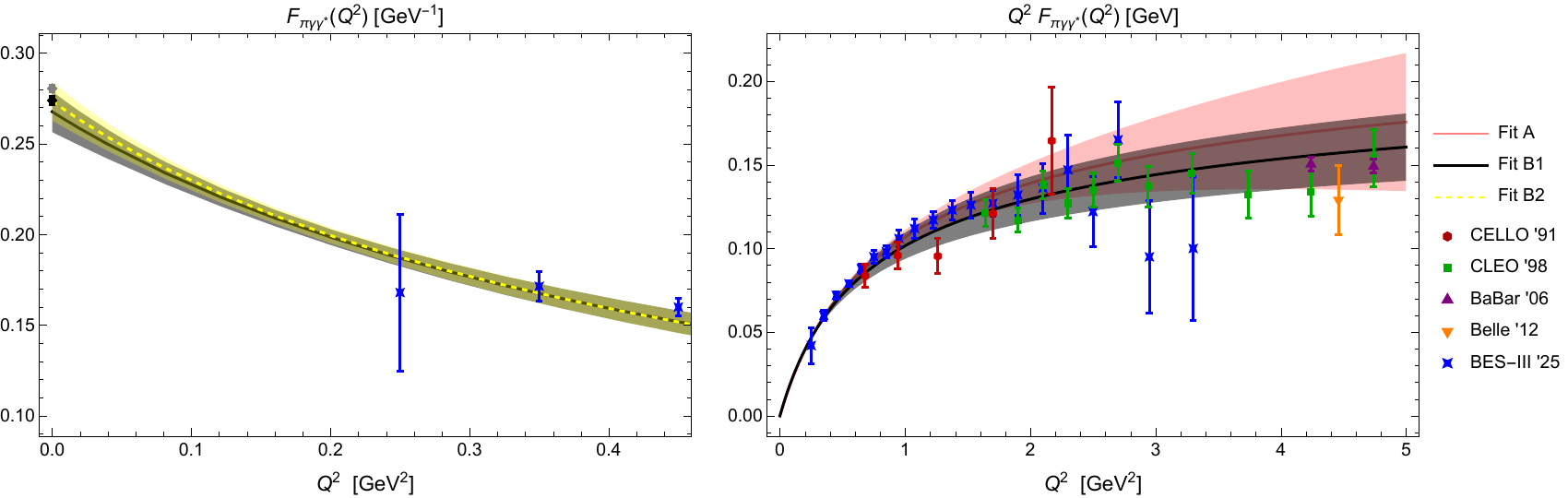}
    \caption{Singly-virtual pion transition form factor in the region of LQCD data. Left panel: Comparison of fits with PDG ({\bf B1}: black solid) and ChPT ({\bf B2}: yellow dashed) normalization constraints. Right panel: Comparison of standalone LQCD ({\bf A}: pink solid) and combined LQCD and experimental ({\bf B1}: black solid) fits.}
    \label{fig:NormPlotB}
\end{figure*}

\begin{figure*}[t]
    \centering
 \includegraphics[width=1.0 \textwidth]{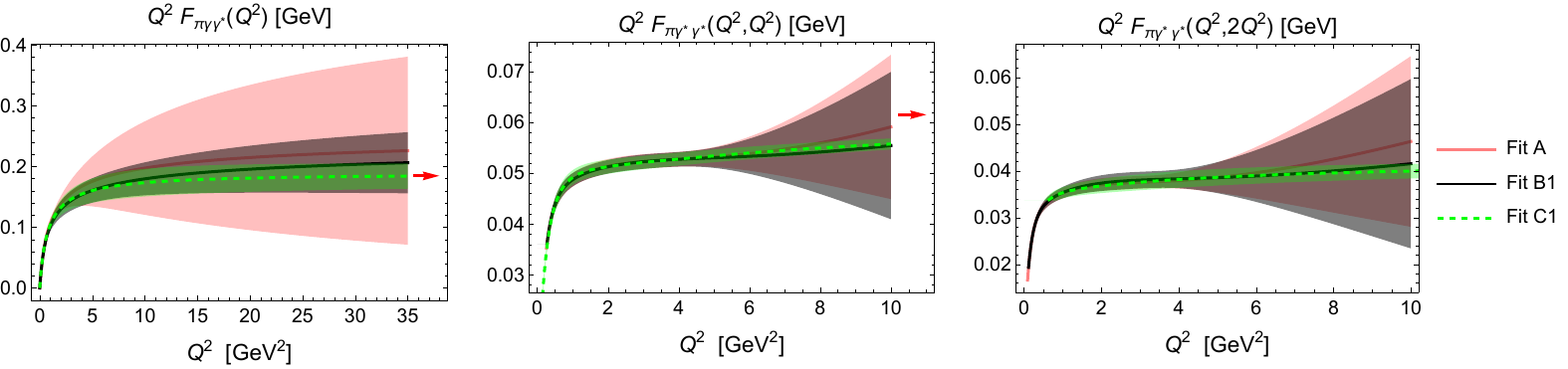}
    \caption{Comparison of fits of the pion transition form factor for different kinematic regions: standalone LQCD ({\bf A}: pink solid), as well as combined LQCD and experimental fits with ({\bf C1}: green dashed) and without ({\bf B1}: black solid) asymptotic constraints. The red arrows indicate the Brodsky-Lepage limit \eref{BL} and the symmetric pQCD limit \eref{OPEsym} for the singly- and doubly-virtual kinematics. }
    \label{fig:plot3panel}
\end{figure*}

As expected, the inclusion of the experimental world data substantially improves the precision of the pion TFF in the singly-virtual photon limit. This improvement is most pronounced in the large-$Q^2$ region shown in \Figref{plot3panel} (left panel), where no LQCD data exist, yielding an uncertainty reduction by more than a factor of three. However, the enhancement is not limited to this kinematic regime; as can be seen in \Figref{NormPlotB} (right panel), the precision also improves by roughly a factor of two at lower virtualities. Incorporating the experimental data causes a downward shift of the central value of the TFF, which is further enhanced when enforcing the asymptotic BL limit. The symmetric OPE limit, indicated in \Figref{NormPlotB} (middle panel), takes effect at slightly larger values of $Q^2$ (note the different axes). For photon virtualities $Q^2>5\,\mathrm{GeV}^2$ in the doubly-virtual kinematics, the TFF is entirely unconstrained by data. Consequently, the uncertainty bands in \Figref{plot3panel} lose their predictive reliability in that region and represent purely model-driven extrapolations. Finally, the effect of the normalization constraints, highlighted in \Figref{NormPlotB} (left panel), is well confined to the low-$Q^2$ region.

\section{Impact on the muon $g-2$} \seclab{gm2}

We now quantify the impact of a combined LQCD and experimental analysis on the muon anomalous magnetic moment  through the pion-pole contribution, $a_\mu^{\pi^0}$, shown in \Figref{pion_pole}. The general master formula for the HLbL contribution to $a_\mu$ \cite{Colangelo:2015ama} simplifies substantially for pseudoscalar-meson poles:
\begin{align}
\eqlab{weighting}
a_\mu^{\pi^0} &= \frac{2\alpha^3}{3\pi^2} \int_0^\infty \dd Q_1 \int_0^\infty \dd Q_2 \int_{-1}^1 \dd \tau \sqrt{1-\tau^2}\, Q_1^3 Q_2^3 \nn\\
\times& \left[  T_1\,\overline{\Pi}_1 + T_2\,\overline{\Pi}_2 \right] (Q_1, Q_2,\tau) \nn
\end{align}
with the HLbL scalar amplitudes
\begin{subequations}
\begin{align}
\overline {\Pi}_1 (Q_1, Q_2, \tau)&= -\frac{\mathcal{F}_{\pi^0\gamma^*\gamma^*}(Q_1^2, Q_3^2)\mathcal{F}_{\pi^0\gamma\gamma^*}(Q_3^2)}{Q_3^2 + M_\pi^2} , \\
\overline{\Pi}_2 (Q_1, Q_2, \tau)&= -\frac{\mathcal{F}_{\pi^0\gamma^*\gamma^*}(Q_1^2, Q_3^2)\mathcal{F}_{\pi^0\gamma\gamma^*}(Q_2^2)}{Q_2^2 + M_\pi^2} ,
\end{align}
\end{subequations}
the QED integral kernels $T_1$ and $T_2$ \cite[Appendix E.2]{Colangelo:2015ama},
and $Q_3^2 = Q_1^2 + 2Q_1Q_2\tau + Q_2^2$.

We evaluate $a_\mu^{\pi^0}$ for the five fits presented in \secref{TFF}. The standalone fit of the LQCD data ({\bf Fit A}) establishes our baseline [in units of $10^{-11}$]:
\begin{subequations}
\begin{eqnarray}
    a_\mu^{\pi^0}(\text{Mainz/CLS}) &=&
    59.7(3.4)_{\text{stat}}(0.9)_{\text{syst}}(0.5)_{\text{disc}},\qquad\eqlab{amuLQCD}\\
    a_\mu^{\pi^0}(\text{\bf Fit A}) &=& 58.4 (3.9)_{\text{stat}}\;.\eqlab{amuThiswork}
\end{eqnarray}
\end{subequations}
Here, we again compare to the original LQCD analysis \cite{Gerardin:2019vio}, whose uncertainty is smaller yet comparable to our prediction. The shift in the central value reflects the difference in normalization discussed in \eref{normcomp}.
Including experimental constraints, our uncertainty for the pion-pole contribution from LQCD alone \eref{amuThiswork} improves by a factor $1.5$ [in units of $10^{-11}$]:
\begin{subequations}
\begin{eqnarray}
    a_\mu^{\pi^0}(\text{\bf Fit B1}) &=& 60.1 (2.7)_{\text{stat}},\\
    a_\mu^{\pi^0}(\text{\bf Fit B2}) &=& 61.0 (2.7)_{\text{stat}},\\
    a_\mu^{\pi^0}(\text{\bf Fit C1}) &=& 61.1  (3.2)_{\text{stat}},\\
        a_\mu^{\pi^0}(\text{\bf Fit C2}) &=& 61.8(2.9)_{\text{stat}}.
\end{eqnarray}
\end{subequations}
Since the weighting of the integral in \Eqref{weighting} is dominated by low-$Q$, the biggest effect is achieved by constraining the normalization of the pion TFF.  This has already been shown in Ref.~\cite{Gerardin:2019vio}, where the normalization of the pion TFF was constrained through the PrimEx experiment \cite{PrimEx:2010fvg}:
\begin{eqnarray}
    a_\mu^{\pi^0}(\text{Mainz/CLS} + \text{PrimEx}) &=& 62.3 (2.3)\times 10^{-11} ,\qquad
\end{eqnarray}
resulting in a comparable improvement over their pure LQCD prediction \eref{amuLQCD}.

\section{Summary and Outlook} \seclab{Summary}

In this work, we presented a combined analysis of LQCD and experimental data for the space-like pion TFF. We combined $24$ LQCD ensembles from the Mainz/CLS Collaboration with singly-virtual measurements from the CELLO, CLEO, BaBar, Belle, and BESIII experiments. To ensure a statistically rigorous combination of datasets with vastly different sizes and unknown experimental correlations, we implemented a global one-stage fitting approach utilizing a normalized $\chi^2$ weighting scheme and synthetic jackknife replicate sampling of the experimental data. We employed the modified $z$-expansion as a flexible model with an appropriate $1/Q^2$ scaling at asymptotic photon virtualities. With this framework, we have successfully established a robust joint fitting methodology that is structurally general and can be directly extended to the $\eta$ and $\eta^\prime$ TFFs.

We have shown that the inclusion of experimental data substantially tightens the constraints on the pion TFF in the singly-virtual kinematics, yielding an uncertainty reduction by a factor of two in the low-$Q^2$ region and more than a factor of three at large $Q^2$. For the pion-pole contribution to the muon anomalous magnetic moment, $a_\mu^{\pi^0}$, this translates to an uncertainty reduction by a factor of $1.5$ as compared to a standalone LQCD baseline. This more modest improvement reflects the fact that the $g-2$ integral is heavily dominated by the low-$Q^2$ region. As expected, the low-$Q^2$ normalization provides the dominant constraint on the final integral.

While lattice evaluations of the pion TFF have reached unprecedented precision, predictions for the heavier pseudoscalars are not yet as advanced as dispersive approaches and remain highly receptive to experimental constraints. Consequently, future extensions of this work will focus on constraining the $\eta$ and $\eta^\prime$ TFFs and refining their contributions to the muon anomalous magnetic moment.

\section*{Acknowledgements}
We would like to thank Antoine G{\'e}rardin, Georg von Hippel, Jonna Koponen, and Harvey B.\ Meyer for fruitful discussions, particularly concerning the underlying lattice methodology, the statistical treatment of the data, and the details of the dataset itself. We are also indebted to them for providing the Mainz/CLS lattice data that made this analysis possible.
This work is supported by the Deutsche Forschungsgemeinschaft (DFG) through the Research Unit FOR 5327 ``Photon-photon interactions in the Standard Model and beyond - exploiting the discovery potential from MESA to the LHC'' (grant 458854507), the Collaborative Research Center 1660 ``Hadrons and Nuclei as Discovery Tools'' (grant 514321794), and
the Emmy Noether Programme (grant 449369623).

\newpage

\appendix
\onecolumngrid

\section{Supplementary material}\seclab{supplement}

\subsection{Best-fit parameters}

Summary of results for {\bf Fits A}, {\bf B1}, and {\bf C1}: best-fit coefficients $c_{nm}$ for the continuum and physical quark masses are listed in Table \ref{tab:zexpansion_parameters},
correlation matrices are given in Eqs.~\eref{corr_matrix_A_aligned}-\eref{corr_matrix_C_aligned}.

\begin{table*}[h]
\centering
\caption{Coefficients of the $z$-expansion in GeV$^{-1}$, defined through \Eqref{zexp} with $N = 3$ and $Q_\mathrm{max}^2=4\,\mathrm{GeV}^2$. }
\label{tab:zexpansion_parameters}
\smallskip
\small
\begin{tabular}{p{1cm}p{1cm} *{10}{r@{.}l}}
\toprule
 & $\chi^2_\text{dof}$&\multicolumn{2}{c}{$c_{00}$} & \multicolumn{2}{c}{$c_{01}$} & \multicolumn{2}{c}{$c_{11}$} & \multicolumn{2}{c}{$c_{20}$} & \multicolumn{2}{c}{$c_{21}$} & \multicolumn{2}{c}{$c_{22}$} & \multicolumn{2}{c}{$c_{30}$} & \multicolumn{2}{c}{$c_{31}$} & \multicolumn{2}{c}{$c_{32}$} & \multicolumn{2}{c}{$c_{33}$} \\
\midrule
\textbf{Fit A} & 1.32 &0&2348(62) & $-0$&0765(55) & $-0$&329(84)  & 0&138(41) & $-0$&008(130) & $-0$&75(62) & 0&37(12) & 0&11(38) & $-1$&07(68) & 0&98(1.25)  \\
\textbf{Fit B1} &1.27& 0&2351(69) & $-0$&0757(63) & $-0$&302(92)  & 0&124(48) & $-0$&127(180) & $-0$&95(76) & 0&42(18) & 0&31(33) & $-1$&05(36) & 0&37(1.27)  \\
\textbf{Fit C1} &1.22 &0&2339(67) & $-0$&0614(95) & $-0$&345(89)  & 0&144(47) & $-0$&186(157) & $-0$&95(76) & 0&27(16) & 0&51(21) & $-0$&48(23) & \multicolumn{2}{c}{}  \\
\bottomrule
\end{tabular}
\end{table*}

\begin{equation}
\text{corr}(\boldsymbol{\mathrm{A}}) = \begin{pmatrix}
+1.000 & -0.076 & -0.366 & +0.137 & -0.081 & -0.158 & +0.056 & +0.315 & -0.048 & -0.293 \\
-0.076 & +1.000 & -0.132 & +0.018 & -0.308 & +0.348 & -0.312 & +0.005 & +0.506 & -0.513 \\
-0.366 & -0.132 & +1.000 & -0.858 & +0.595 & +0.312 & -0.569 & -0.396 & +0.115 & +0.383 \\
+0.137 & +0.018 & -0.858 & +1.000 & -0.705 & -0.383 & +0.794 & +0.080 & -0.022 & -0.147 \\
-0.081 & -0.308 & +0.595 & -0.705 & +1.000 & +0.001 & -0.709 & +0.072 & -0.403 & +0.482 \\
-0.158 & +0.348 & +0.312 & -0.383 & +0.001 & +1.000 & -0.327 & -0.637 & +0.496 & -0.014 \\
+0.056 & -0.312 & -0.569 & +0.794 & -0.709 & -0.327 & +1.000 & -0.108 & -0.116 & +0.064 \\
+0.315 & +0.005 & -0.396 & +0.080 & +0.072 & -0.637 & -0.108 & +1.000 & -0.514 & -0.386 \\
-0.048 & +0.506 & +0.115 & -0.022 & -0.403 & +0.496 & -0.116 & -0.514 & +1.000 & -0.534 \\
-0.293 & -0.513 & +0.383 & -0.147 & +0.482 & -0.014 & +0.064 & -0.386 & -0.534 & +1.000
\end{pmatrix}
\label{eq:corr_matrix_A_aligned}
\end{equation}

\begin{equation}
\text{corr}(\boldsymbol{\mathrm{B1}}) = \begin{pmatrix}
+1.000 & -0.435 & +0.624 & -0.293 & -0.420 & -0.394 & +0.487 & +0.376 & +0.039 & -0.481 \\
-0.435 & +1.000 & -0.356 & +0.332 & +0.267 & +0.244 & -0.544 & +0.136 & +0.418 & -0.117 \\
+0.624 & -0.356 & +1.000 & -0.909 & +0.008 & +0.320 & +0.043 & -0.551 & +0.289 & +0.224 \\
-0.293 & +0.332 & -0.909 & +1.000 & +0.018 & -0.332 & -0.020 & +0.378 & -0.506 & +0.094 \\
-0.420 & +0.267 & +0.008 & +0.018 & +1.000 & +0.800 & -0.888 & -0.664 & +0.153 & +0.241 \\
-0.394 & +0.244 & +0.320 & -0.332 & +0.800 & +1.000 & -0.815 & -0.400 & +0.458 & +0.411 \\
+0.487 & -0.544 & +0.043 & -0.020 & -0.888 & -0.815 & +1.000 & +0.498 & -0.468 & -0.306 \\
+0.376 & +0.136 & -0.551 & +0.378 & -0.664 & -0.400 & +0.498 & +1.000 & -0.101 & -0.780 \\
+0.039 & +0.418 & +0.289 & -0.506 & +0.153 & +0.458 & -0.468 & -0.101 & +1.000 & -0.415 \\
-0.481 & -0.117 & +0.224 & +0.094 & +0.241 & +0.411 & -0.306 & -0.780 & -0.415 & +1.000
\end{pmatrix}
\label{eq:corr_matrix_B_aligned}
\end{equation}

\begin{equation}
\text{corr}(\boldsymbol{\mathrm{C1}}) = \begin{pmatrix}
+1.000 & -0.190 & +0.134 & -0.247 & -0.332 & -0.375 & +0.414 & +0.318 & -0.374 \\
-0.190 & +1.000 & -0.074 & -0.067 & -0.096 & +0.645 & -0.478 & -0.291 & +0.447  \\
+0.134 & -0.074 & +1.000 & -0.941 & +0.128 & +0.283 & -0.157 & -0.491 & +0.309  \\
-0.247 & -0.067 & -0.941 & +1.000 & -0.110 & -0.351 & +0.193 & +0.497 & -0.333 \\
-0.332 & -0.096 & +0.128 & -0.110 & +1.000 & +0.617 & -0.805 & -0.822 & +0.797 \\
-0.375 & +0.645 & +0.283 & -0.351 & +0.617 & +1.000 & -0.954 & -0.902 & +0.961  \\
+0.414 & -0.478 & -0.157 & +0.193 & -0.805 & -0.954 & +1.000 & +0.928 & -0.987  \\
+0.318 & -0.291 & -0.491 & +0.497 & -0.822 & -0.902 & +0.928 & +1.000 & -0.972  \\
-0.374 & +0.447 & +0.309 & -0.333 & +0.797 & +0.961 & -0.987 & -0.972 & +1.000  \\
\end{pmatrix}
\label{eq:corr_matrix_C_aligned}
\end{equation}

\newpage

\subsection{Jackknife samples of experimental data}

\begin{figure*}[htb]
    \centering
 \includegraphics[width=0.4 \textwidth]{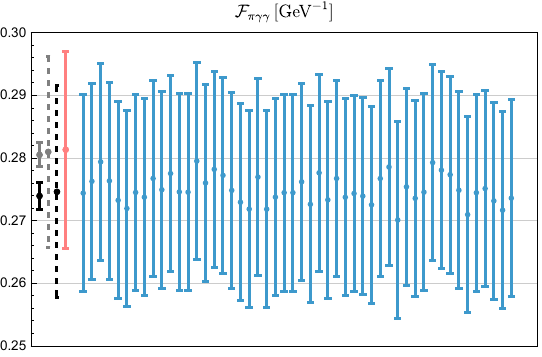}
    \caption{Visualization of the synthetic jackknife replicates of the experimental data $\mathcal{F}_{\pi\gamma\gamma^*}(Q^2)$: $N_\text{Jk}=50$ synthetic jackknife samples (blue), jackknife mean (pink), experimental fit with PDG normalization \eref{normalizationconstraintPDG} (black dashed), experimental fit with ChPT normalization \eref{normalizationconstraintChPT} (gray dashed), experimental data points are the same as in \Figref{FitPlotPionTFF}. 
    See \secref{synthetic} for details. }
    \label{fig:ExperimentalJKsamplesNormPlot}
\end{figure*}
\begin{figure*}[htb]
    \centering
 \includegraphics[width=0.85\textwidth]{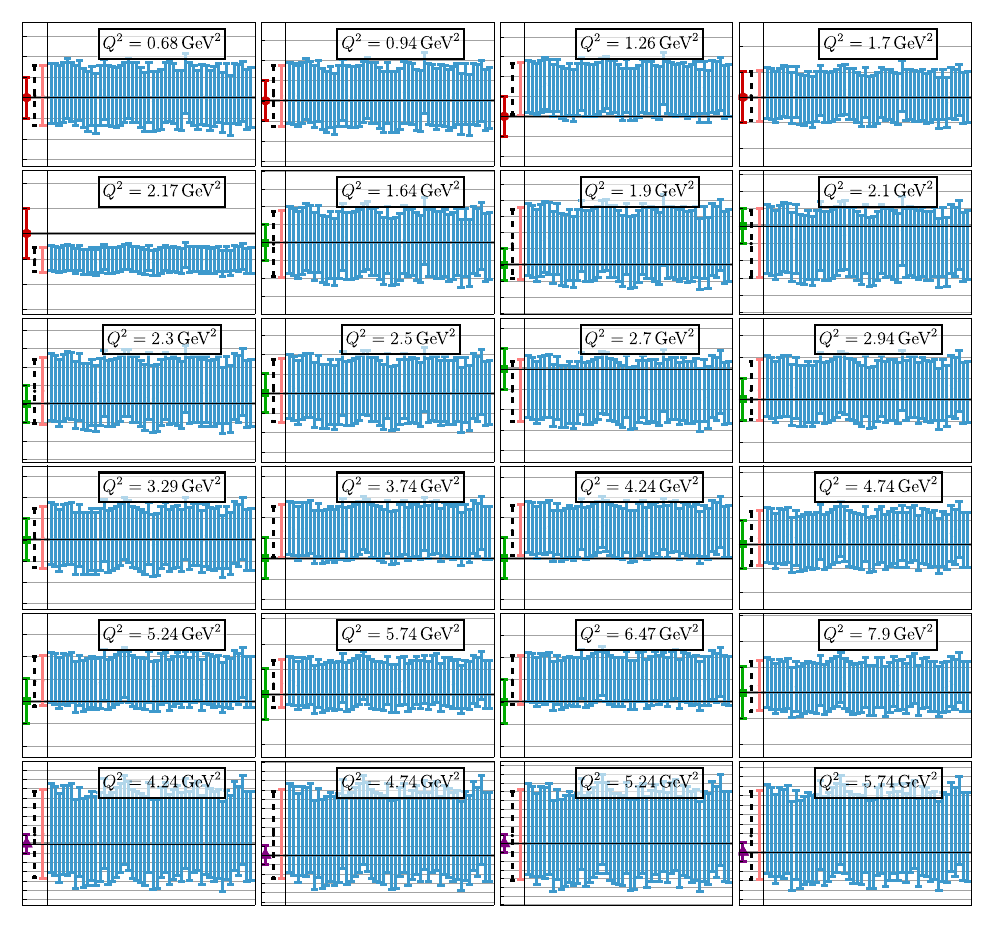}
    \caption{Legend is the same as in Fig.~\ref{fig:ExperimentalJKsamplesNormPlot}. Grid lines indicate $1\,\sigma$ intervals of the experimental data points.} \label{fig:ExperimentalJKsamplesNormalization_1}
\end{figure*}
\begin{figure*}[htb]
    \centering
 \includegraphics[width=0.85 \textwidth]{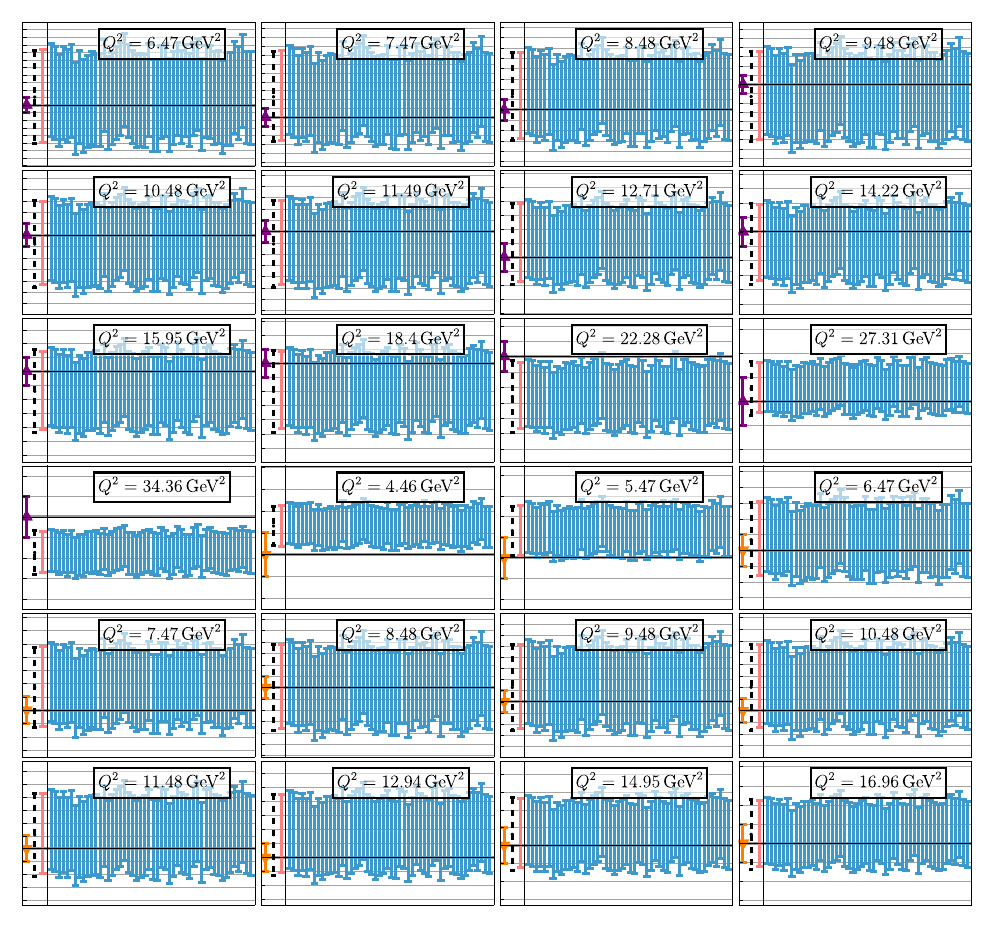}
    \caption{Legend is the same as in Fig.~\ref{fig:ExperimentalJKsamplesNormPlot}. Grid lines indicate $1\,\sigma$ intervals of the experimental data points.}  \label{fig:ExperimentalJKsamplesNormalization_2}
\end{figure*}
\begin{figure*}[htb]
    \centering
 \includegraphics[width=0.85 \textwidth]{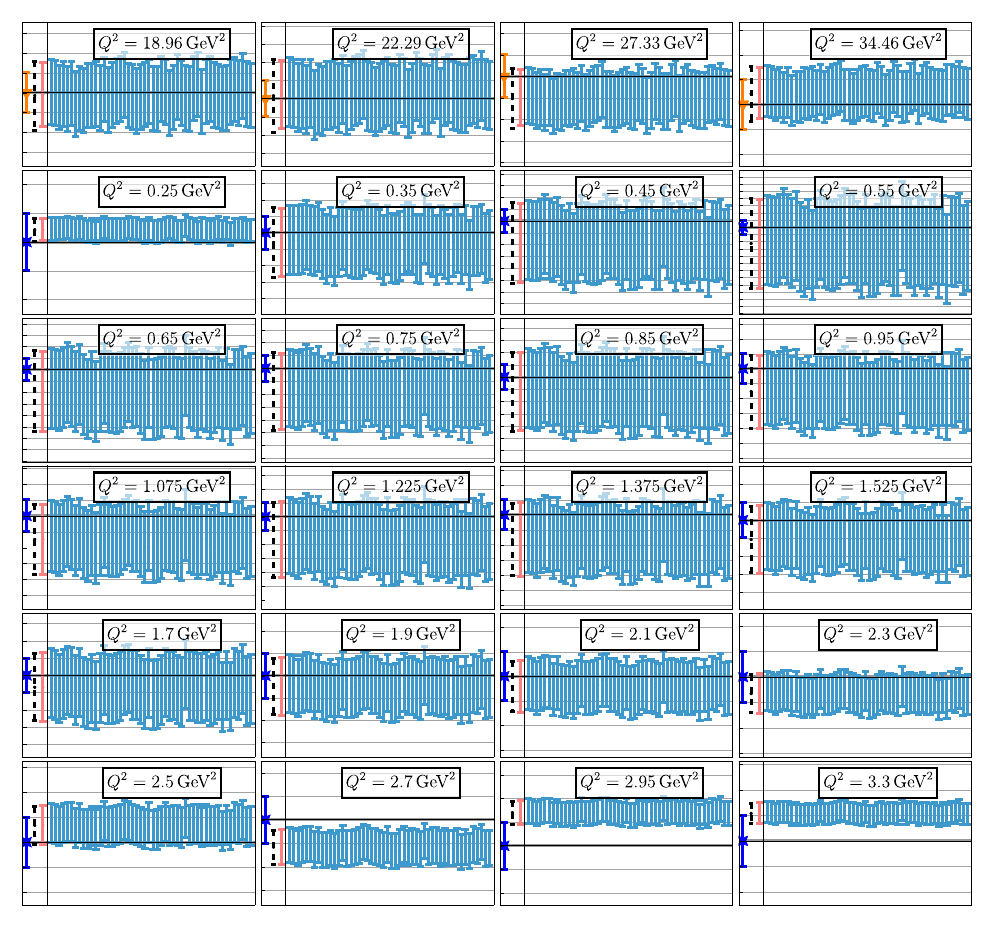}
    \caption{Legend is the same as in Fig.~\ref{fig:ExperimentalJKsamplesNormPlot}. Grid lines indicate $1\,\sigma$ intervals of the experimental data points.}
    \label{fig:ExperimentalJKsamplesNormalization_3}
\end{figure*}

\end{document}